\documentclass[11pt]{article}
\usepackage{fullpage}
\usepackage{amsmath}
\usepackage{amssymb}
\usepackage{setspace}
\usepackage{bbm}
\usepackage{dsfont}
\usepackage{graphics}
\usepackage{amsmath}
\usepackage{enumerate}
\usepackage{amsfonts}
\usepackage[all]{xy}
\usepackage[colorlinks=true,linktocpage=true,linkcolor=blue,citecolor=blue]{hyperref}
\usepackage[numbers,sort&compress]{natbib}
\usepackage{graphicx}
\usepackage{amsmath,graphicx,amssymb,subfigure}
\usepackage{amssymb}
\usepackage[section]{placeins}
\newcommand{\bep}{\boldsymbol{\varepsilon}}

\usepackage{color}

\usepackage[colorlinks=true]{hyperref} 
\hypersetup{
    bookmarks=true,         % show bookmarks bar?
    unicode=false,          % non-Latin characters 
    pdftoolbar=true,        % show Acrobat
    pdfmenubar=true,        % show Acrobat 
    pdffitwindow=false,     % window fit to page when opened
    pdfstartview={FitH},    % fits the width of the page to the window
    pdftitle={My title},    % title
    pdfauthor={Author},     % author
    pdfsubject={Subject},   % subject of the document
    pdfcreator={Creator},   % creator of the document
    pdfproducer={Producer}, % producer of the document
    pdfkeywords={keyword1} {key2} {key3}, % list of keywords
    pdfnewwindow=true,      % links in new window
    colorlinks=true,       % false: boxed links; true: colored links
    linkcolor=blue,          % color of internal links (change box color with linkbordercolor)
    citecolor=red,        % color of links to bibliography
    filecolor=magenta,      % color of file links
    urlcolor=cyan           % color of external links
}

\usepackage[normalem]{ulem}
\usepackage{amsmath}
\usepackage{enumerate}
\usepackage{amsfonts}
\usepackage{yfonts}

\usepackage{subfigure}
\usepackage{psfrag}

\usepackage{epsfig}
\usepackage[latin1]{inputenc}
\usepackage{float}
\usepackage{graphicx}
\usepackage{cancel}
\usepackage{mathrsfs}
\usepackage{amssymb}
\usepackage{amsfonts}
\usepackage{amsmath}
\usepackage{slashed}
\usepackage[font=small,labelfont=bf]{caption}

\newcommand \tr {\mbox{{\bf Tr}}}

\usepackage{graphicx}
\usepackage{bm}

\newcommand{\be}{\begin{equation}}
\newcommand{\ee}{\end{equation}}
\newcommand{\bes}{\begin{equation*}}
\newcommand{\ees}{\end{equation*}}
\newcommand{\bea}{\begin{eqnarray}}
\newcommand{\eea}{\end{eqnarray}}
\newcommand{\beas}{\begin{eqnarray*}}
\newcommand{\eeas}{\end{eqnarray*}}

\newcommand{\bmat}{\begin{bmatrix}}
\newcommand{\emat}{\end{bmatrix}}

\newcommand{\WE}{\wedge}
\newcommand{\D}{\delta}
\newcommand{\ep}{\epsilon}
\newcommand{\de}{\nabla}

\begin{document}

\numberwithin{equation}{section}
{
\begin{titlepage}
\begin{center}

\hfill \\
\hfill \\
\vskip 0.75in

{\Large \bf Relative Entropy, Mixed Gauge-Gravitational Anomaly and Causality}\\

\vskip 0.4in

{\large Arpan Bhattacharyya${}^{a,c}$, Long Cheng${}^{a}$ and Ling-Yan Hung${}^{a,b}$}

\vskip 0.3in

{\it ${}^{a}$ Department of Physics and Center for Field Theory and Particle Physics, Fudan University, \\
220 Handan Road, 200433 Shanghai, China} \vskip .5mm
{\it ${}^{b}$ Collaborative Innovation Center of Advanced  Microstructures,
Fudan University,\\ 220 Handan Road,  200433 Shanghai, China.}\vskip.5mm
{\it ${}^{c}$Centre For High Energy Phsyics, Indian Institute of Science, 560012 Bangalore, India}

\end{center}

\vskip 0.35in

\begin{center} {\bf ABSTRACT }
 \end{center}
In this note we explored the holographic relative entropy in the presence of the 5d Chern-Simons term, which introduces a mixed gauge-gravity anomaly to the dual CFT. The theory trivially satisfies an entanglement first law. However, to quadratic order in perturbations of the stress tensor $T$ and current density $J$, there is a mixed contribution to the relative entropy bi-linear in $T$ and $J$, signalling a potential violation of the positivity of the relative entropy. Miraculously, the term vanishes up to linear order in a derivative expansion. This prompted a closer inspection on a different consistency check, that involves time-delay of a graviton propagating in a charged background, scattered via a coupling supplied by the Chern-Simons term. The analysis suggests that the time-delay can take either sign, potentially violating causality for any finite value of the CS coupling.

\vfill

\noindent \today

\end{titlepage}
}
%%%%%%%%%%%%%%%%%%%%%%%%%%%%%%%%%%%%%%%%%%%%%%%%%%%%%%%%%%%%%%%%%%

%%%%%%%%%%%%%%%%%%%%%
\newpage

\tableofcontents
%\newpage
\setlength\arraycolsep{1pt}

\section{Introduction}

It is growingly clear that many body entanglement holds important information about the structure of field theory, and in which one could uncover deep insights about gravity and perhaps quantum gravity via the gauge gravity correspondence. The study of holographic entanglement entropy alone for example, has led to many important new understanding. One important line of study regarding the relative entropy led to insights on the connection between the first law of entanglement and the equations of motion of a bulk gravity theory \cite{Lashkari, MyersHartman, Swingle}.  As one moves on to study perturbations at higher orders, it is realized that the relative entropy detects hidden pathology of a theory \cite{Me}, and recovers many constraints on the couplings of higher derivative terms  that were found in prior study using other arguments based for example on causality \cite{Hofman, Hofman1,Brigante}.

Only relatively recently, there is a surge of interest in understanding entanglement entropy also of anomalous systems, partly propelled by their connection to many condensed matter systems such as the boundary states of symmetry protected topological (SPT) phases and other topological orders \cite{Wen,Wen1,Wen2,Wen3}. It is thus also naturally of interest to understand their manifestation in a dual gravity theory. The dual theories generically comprise of Chern-Simons terms, and their contribution to the bulk entanglement entropy functional were found in a series of works \cite{Castro, Guo, Gimseng, Nishioka}.  The physics encoded in the entanglement entropy of anomalous systems is only systematically studied very recently \cite{Iqbal}. And there are also some works studying the implications of these anomalous contributions in the gravity dual. For example, it is found that the first law of entanglement is satisfied in topologically massive gravity (TMG) \cite{Janet1} despite the appearance of new modes dual to non-unitary operators that source the metric linearly, which usually signals a possible violation of unitarity. As demonstrated in \cite{Janet, Wong1}, a quadratic coupling of the matter sector to gravity was crucial towards recovering a positive definite relative entropy.   In this paper, we study in detail the 5d Chern-Simons term, which is responsible for a mixed gauge-diffeomorphism anomaly in the dual field theory. This term carries a linear coupling of the gauge field to the gravity sector, and as such raises suspicion for the same reasoning alluded to above that positivity of the relative entropy could be violated. We find in this paper that while the first law of entanglement entropy is satisfied, as we begin probing quadratic contribution, there is a coupled contribution of the stress tensor and the current density. Such terms are doomed not to come with a definite sign, and that it can be shown that they do not arise from a perfect square. This term miraculously vanishes to leading order in a derivative expansion.  Nonetheless, it remains plausible that the Chern-Simons coupling should remain only an infinitesimal parameter in order that the dual theory remains unitary. We complement our study  making use of causality.  Following considerations in \cite{Edelstein}  we study propagation of gravitons in a charged background. In the current case, there appears to be a time delay that arises from gravitons scattering off photons instead of other gravitons, via a three point vertex supplied by the CS term. Such a time-delay is found also to take either sign, and could potentially violate causality for any finite value of the CS coupling. 

The paper is organized as follows. In section \ref{sec2} we begin with a check of the first law of entanglement in the presence of the CS term. Then in section \ref{sec3}, we explore a set of corrections to the entanglement entropy that contains $T.J$. It is evident that these terms could potentially violate the positivity of the relative entropy, despite vanishing miraculously at precisely first order derivative level. In section \ref{sec4}, we explore the possibility of cancelling these offending contributions by adding extra terms to the entanglement entropy functional. In section \ref{sec5} we compute time delay of a propagating graviton in a charged background, corroborating the conclusion reached from the discussion of relative entropy.
Finally we end with a conclusion. Some details of the Wald charge and the entanglement entropy functional of the CS term is relegated to the appendix. 

\section{Entanglement First Law of Chern-Simons gravity with $U(1)$ gauge field } \label{sec2}

Given two (reduced) density matrices $\rho,\sigma$, the relative entropy is defined as
\be
S(\rho|\sigma) = \tr(\rho \ln \rho ) - \tr(\rho \ln \sigma).
\ee
This quantity can be demonstrated to be greater than or equal to zero, making use of the positivity of the eigenvalues of the density matrices, that they are properly normalized and that the log is a convex function. Since these matrices are positive definite, one can write 
\be
\sigma = \frac{\exp(-H)}{\tr \exp(-H)},
\ee
for some Hermitian operator $H$. 
The relative entropy can thus be expressed as
\be
S(\rho|\sigma) = \Delta \langle H\rangle - \Delta S, 
\ee 
where $\Delta S = S(\rho) - S(\sigma)$, and $\Delta \langle H\rangle = \tr(\rho H) - \tr (\sigma H)$.

Suppose $\rho = \sigma + \epsilon \delta\rho$, for some infinitesimal deviation $\epsilon \delta \rho$. Then the positivity of the relative entropy would require that there is no linear order dependence on $\epsilon$ when we expand the quantities in $\epsilon$. This is the analogue of the first law, or the entanglement first law for $\rho,\sigma$ reduced density matrices:
\be
\delta S = \delta \langle H \rangle.
\ee
Quadratic corrections in $\epsilon^2$ would then ensure that $\Delta S < \Delta \langle H\rangle$.

To begin with, we would like to check the entanglement first law. 
If the reference state $\sigma$ is the reduced density matrix of a region bounded by a spherical surface in the ground state of a CFT, the entanglement Hamiltonian takes the well known form 
\be \label{modH}
H = {2\pi}\int_{r<R} d^{d-1}x \frac{R^2- r^2}{2R} T_{00}.
\ee

We consider the following action which consists of the usual Einstein-Maxwell action plus five-dimensional Chern-Simons term,  
\begin{align}
\begin{split}  \label{CSL}
S= & S_\textrm{EH}+S_\textrm{maxwell}+S_\textrm{CS}\\&=\int d^{5} x \sqrt{-g}\, \mathcal{L}=\frac{1}{2 \ell_{p}^3}\Big[\int \sqrt{g}\, d^{5} x \left( (R-2\Lambda)*1-\frac{1}{2}F\wedge*F+\lambda_{cs}\int\tr(A\WE R\WE R)\right)\Big],
\end{split}
\end{align}
where $2\ell_{p}^3 = 16 \pi G_5$ is the five-dimensional gravitational coupling constant and two-form $F=dA$ is the field strength of the $U(1)$ gauge field $A$ and $g$ denotes the determinant of the bulk metric.\footnote{ if one wants to match with the convention found in the literature \cite{MyersSinhaPaulos, Landsteiner}  one has to replace the $\lambda_{cs}$ in our notation by $4\lambda_{cs}$.} 
The entanglement entropy functional in the presence of  generic Chern-Simons terms has been obtained in \cite{Gimseng}. We will recover below the explicit form for the particular action (\ref{CSL}). To check the first law of entanglement for a spherical entangling surface with respect to the ground state of the boundary CFT, it requires perturbing the holographic entanglement entropy around the pure AdS background.  It has been demonstrated in a rather generic setting that the entanglement entropy would coincide with the Wald charge \cite{WaldIyer} evaluated on the minimal bulk entangling surface, which coincides with a killing horizon, at least to linear order perturbation around the pure AdS background \cite{Lashkari, MyersHartman, Swingle}. 
Since we are working at linear order perturbation, perturbation to the entangling surface itself does not contribute at this order, given that the zeroth-order entangling surface is a solution to the equations of motion following from the entanglement entropy functional. This is also observed in \cite{Janet}. The derivation of the Wald charge for the Chern-Simons theory is detailed in the appendix. The result is given by $Q=\frac{1}{2}Q^{ag}\bep_{ag}$, where
\bea  
Q^{ag}_{CS}=\ep^{a gbde}A_bR_{de}^{~~fv}\de_f\xi_v-\frac{\ep^{a fbde}}{2}F_{fb}R_{de}^{~~gv}\xi_v+\frac{\ep^{gbcde}}{2}F_{bc}R_{de}^{~~a f}\xi_f-\frac{\ep^{fbcde}}{2}F_{fc}R_{de}^{~~a g}\xi_b.\nonumber\\
\eea
and the total Wald charge is given by
\be \label{waldcharge}
Q_{tot}^{ag}= \frac{1}{2\ell_{p}^3}\Big(Q_{EH}^{ag}+Q_{maxwell}^{ag}+\lambda_{cs} Q_{CS}^{ag}\Big),
\ee
where  
\be
Q_{EM}^{ag} = -2\epsilon^{agcd}\nabla_{c}\xi_{d}\qquad Q_{maxwell}^{ag}= F^{a g}\xi^bA_b.
\ee
We have used $a,b,c,\cdots$ etc for denoting the bulk five dimensional indices. $\epsilon^{abcde}$ is the Levi-Civita symbol and we set $\epsilon^{ztxyw}=1.$ Then integrating $Q$ over the bifurcation surface ($\Sigma$) we will get the Wald entropy for this  theory.
\be \label{Waldentropy}
S_{Wald}=-2\pi  \int_{\Sigma} d^{3} x \sqrt{h} \frac{\partial \mathcal{L}}{\partial R_{abcd}}\hat \epsilon^{ab}\hat \epsilon^{cd}=S_{Wald, cs}+S_{Wald,Einstein},
\ee
where,
$h$ is the determinant of the three dimensional induced metric on the codimension 2 bifurcation surface and $\hat \epsilon^{ab}$ is the binormal. Also for our case the Lagrangian $\mathcal{L}$ is defined in (\ref{CSL}) 
\be \label{Waldnetropy1}
S_{Wald, Einstein}=\frac{2\pi}{\ell_{p}^3}\int_{\Sigma} d^{3} x \sqrt{h}\,,\quad S_{Wald, CS}=-\frac{\pi \lambda_{cs}}{2\ell_{p}^3}\int_{\Sigma} d^{3} x\sqrt{h}\, \tilde \epsilon^{abcde}A_{a}R_{bc}{}^{fg}\hat \epsilon_{de}\hat \epsilon_{fg},
\ee
where the Levi-Civita tensor $\tilde \epsilon^{abcde}=\frac{\epsilon^{abcde}}{\sqrt{-g}}.$  To obtain the local expression for entropy (\ref{CSL})  from the Noether charge (\ref{waldcharge}) we have followed the same procedure as detailed in  \cite{WaldIyer}.  For a stationary bifurcation killing surface we have $\nabla_{a}\xi_{b}=\kappa \hat \epsilon_{ab}.$ Further $\xi^{b}$ vanishes everywhere on the bifurcation killing surface.  As explained in \cite{WaldIyer} all the potential ambiguity terms that can enter in the entropy expression vanish on the bifurcation killing surface.\footnote{ Also for this reason only $\nabla_{f}\xi_{v}$ term in (\ref{waldcharge}) contributes to the Wald entropy expression (\ref{Waldnetropy1}).}   Using these two facts \cite{WaldIyer} we can recover the local expression for the Wald entropy (\ref{Waldnetropy1}) from the Wald charge (\ref{waldcharge}) which is valid only on the bifurcation surface. 

\subsection{Linearized Fefferman-Graham expansion}
To check the first law, we consider a small charge density $J$ and stress tensor $T$ in an excited state close to the ground state. To linear order in $J$ and $T$, the correction to the background metric is completely captured by the Fefferman-Graham expansion. 
The corresponding metric and gauge field is thus given by
\be
ds^2=\frac{1}{z^2}\Big ( {dz^2}+ (\eta_{\mu\nu} +h_{\mu\nu}) dx^{\mu} dx^{\nu} \Big),
\ee
where $h_{\mu\nu}$ is the small deviation around the pure AdS background and we will set the AdS radius $L_{AdS}=1$ throughout the paper, whereas
\be
A_{\mu}= A_{\mu}+ \delta A_{\mu}.
\ee
$A_\mu$ is itself vanishing in the pure AdS background. 
First, we consider constant $T$ and $J$. In which case, $h_{\mu\nu}$ admits the following FG expansion \cite{Skenderis,Clark},
\be
h_{\mu\nu}(z, x_{i} )= z^2 h^{(2)}_{\mu\nu}(x_{\mu})+ z^4 h^{(4)}_{\mu\nu}(x_{\mu}) + z^4 \log(z^2)\, t_{
\mu\nu},
\ee
and
\be
\delta A_{\mu}(\rho,x_{i})=\delta A^{(0)}_{\mu}(x_{i})+z^2 \delta A^{(2)}_{\mu}(x_{i})+ z^2 \log(z^2)\, \delta B_{\mu}(x_{i}).
\ee
We have used Greek letters to denote the indices for the four dimensional boundary and $i,j,\cdots$ etc to denote the three spatial indices. We will choose a gauge condition for $A_{\mu}$ such that $\delta A_{z}=0$. 
By solving the equation of motion order by order in the perturbation one can find that \cite{Clark},
\be
h^{(2)}_{\mu\nu}=\frac{1}{2} \Big[R^{(0)}_{\mu\nu}-\frac{1}{6}g^{(0)}_{\mu\nu}R^{(0)}\Big],
\ee
where $g^{(0)}_{\mu\nu}$ is the boundary metric for the background which in our case is $\eta_{\mu\nu}.$ So $h^{(2)}_{\mu\nu}=0.$ 
\be
g^{(0) \mu\nu}h^{(4)}_{\mu\nu}=\frac{1}{16}\Big(R^{(0)}_{\mu\nu}R^{(0)\mu\nu}-\frac{2}{9}R^{(0)}{}^2\Big)  -\frac{\ell_{p}^6}{12}\delta F^{(0)}_{\mu\nu}\delta F^{(0)\mu\nu}, 
\ee
where, $\delta F^{(0)}_{\mu\nu}=\partial_{[\mu} \delta A^{(0)}_{\nu]}$,
which is quadratic in $\delta A$, and thus dropped to the order we are working with here. 
Also following \cite{Clark},
\begin{align}
\begin{split}
t_{\mu\nu}=&\frac{1}{8}\Big[R^{(0)}_{\mu\rho}R^{(0)\rho}_{\nu}-\frac{1}{3}R^{(0)}R^{(0)}_{\mu\nu}-\frac{1}{36}g^{(0)}_{\mu\nu}R^{(0)}{}^2\Big]-\frac{1}{32}g^{(0)}_{\mu\nu}\Big[R^{(0)}_{\rho\sigma}R^{(0)\rho\sigma}-\frac{2}{9}R^{(0)}{}^2 \Big] \\&-\frac{1}{8}\Big[R^{(0)}_{\mu\rho}R^{(0)\rho}_{\nu}-R^{(0)\rho\sigma}R^{(0)}_{\sigma\mu\nu\rho}+\frac{1}{6}\nabla_{\mu}^{(0)}\nabla_{\nu}^{(0)}R^{(0)}-\frac{1}{2}\nabla^{(0)}_{\rho}\nabla^{(0)\rho}R^{(0)}_{\mu\nu}+\frac{1}{12}g^{(0)}_{\mu\nu}\nabla_{\rho}^{(0)}\nabla^{(0)\rho}R^{(0)}\Big]\\&-\frac{\ell_{p}^6}{2}\Big[\delta F^{(0)}_{\mu\rho}\delta F^{(0)\rho}_{\nu}+\frac{1}{4} g^{(0)}_{\mu\nu}\delta F^{(0)}_{\rho\sigma}F^{(0)\rho\sigma}\Big].
\end{split}
\end{align}
So,        
\be
g^{(0)\mu\nu} t_{\mu\nu}=0 
\ee
and
\be
\delta B^{\mu}=\frac{1}{4}\partial_{\nu}\delta F^{(0) \nu\mu}.
\ee
As reviewed briefly  above, the entanglement entropy is the surface integral of the conserved Wald charge (\ref{waldcharge}) evaluated at the bifurcation surface. By charge conservation therefore this is equal to the surface integral of the charge in the boundary of AdS, such that together with the bifurcation surface these two surfaces form a closed surface. Therefore the integral at the bifurcation surface is equal to that in the boundary AdS surface. This equality is the basis of the first law satisfied by black holes \cite{WaldIyer}. To demonstrate the entanglement first law, we need to demonstrate that the conserved Wald charge evaluated at the AdS boundary, i.e. $\int_{z=0, t=t_{0}} \delta Q$, in fact coincides with the change of the expectation value of the entanglement Hamiltonian itself. 

Substituting the Fefferman-Graham expansion into the linearized expression for $\delta Q_{CS}$, and recall that the minimal surface is a killing horizon, whose only non-zero component at $z=0$ is
\be
\zeta^{t}=\frac{2\pi}{R} (R^2-|x^{i}-x_{0}^{i}|^2),
\ee
we get
\be
\int_{z=0, t=t_{0}} \delta Q_{CS}=0.
\ee
The Chern Simons term does not contribute to the asymptotic energy at the linearized order in perturbation. With the current form of the Fefferman Graham asymptotic expansion of the metric, it was demonstrated already in \cite{Janet1, Janet, Lashkari,MyersHartman} that the entanglement first law is in fact satisfied. i.e. Without the CS part, the Einstein-Maxwell part of the Wald charge evaluated on this same metric perturbation already recovers the correct expectation value of the entanglement Hamiltonian of the spherical surface. Therefore the first law would only continue to hold if the CS terms in fact makes no contribution at this order, which, fortunately, is indeed the case.  Note that as we are perturbing around AdS vaccum, we have set $\delta A^{(0)}_{\mu}=0$.

We note that $\delta Q_{CS}=0$ can be expected also from the following, namely that
$\delta R_{ab}{}^{cd}$ near $z=0$ and $t=t_{0}$ start  from $\mathcal{O}(z^4).$ Also  $\nabla_{a}\zeta_{b}$  at the zeroth order starts from $\mathcal{O}(\frac{1}{z^2}).$ It is thus evident that $\delta Q_{CS}$ around $z=0$ will not have any non zero contribution as $z\rightarrow 0.$

In \cite{MyersHartman}, it was demonstrated on very general grounds that the entanglement first law for a spherical region in holographic duals of CFTs containing higher derivative terms is satisfied when there is bulk diffeomorphism invariance, even if the theory should fail to preserve unitarity. In the current case, since the CS term considered preserve bulk diffeomorphism, it is indeed expected that the first law should continue to hold. Nonetheless, it is reassuring to see how it emerges here explicitly, as we set the stage and notation for explorations of higher order terms  in the next sections that exhibit some special features pertaining to the unique nature of the Chern-Simons terms. 

\section{Mixed correction-- violation of positivity?} \label{sec3}
One important feature of the Chern-Simons term is a linear coupling of the gauge field with gravity. As emphasized in the introduction, quadratic coupling of generic matter fields with gravity ensured positivity of the relative entropy at quadratic order of the perturbation.  This is no longer guaranteed in the Chern-Simons theory. 
To check that, we revisit the correction to the entanglement entropy, now computing to order quadratic in $T$ and $J$. In particular, any appearance of $T.J$ coupled terms would immediately signal a potential violation of the positivity of the relative entropy. And this is going to be the focus of the paper. 

As it will become clear, any potential coupling term between $J$ and $T$ has any hope of being non-vanishing if they contain spatial and temporal dependence. We will assume that the modulation is very small, and thus admit a derivative expansion. The Fefferman Graham expansion, to leading order in the derivative expansion, takes the following form:  
\be
\delta A_{\mu}= \,z^2 J_{\mu} +z^6  T_{\mu\nu}J^{\nu},
\ee
where
\begin{align}
\begin{split} \label{pert1}
h_{\mu\nu}= &z^4 T_{\mu\nu}+  z^6 (\tilde \alpha_{1} J_{\mu \kappa}J^{\kappa}_{\nu}+\tilde \alpha_{2}\eta_{\mu\nu} J^2) + \tilde \alpha \, z^8 \epsilon^{\kappa\lambda\delta}{}_{(\mu} J_{\kappa}\partial_{\delta}T_{\lambda |\nu)}+  \tilde\beta\, z^8 \epsilon^{\kappa\lambda\delta}{}_{(\mu} \partial_{\delta}J_{\kappa}T_{\lambda |\nu)}\\&+ z^8 (
\tilde \beta_{1} T_{\mu\sigma}T^{\sigma}_{\nu}+\tilde \beta_{2} \eta_{\mu\nu} T^2)+\cdots,
\end{split}
\end{align}
and that $\tilde\alpha_{1}, \tilde\alpha_{2}, \tilde\alpha, \tilde\beta, \tilde\beta_{1},\tilde \beta_{2}$ can be found by solving equations of motion. 

In particular, we have \footnote{$\tilde\alpha_{1}$ and $\tilde\alpha_{2}$ have been already calculated in \cite{Janet}.}
\be\tilde \alpha_{1}=-\frac{1}{12},\qquad \tilde\alpha_{2}= \frac{1}{72},\qquad \tilde \alpha=-2\lambda_{cs},\qquad \tilde \beta=3\lambda_{cs},\qquad \tilde \beta_{1}=\frac{1}{2},\qquad \tilde \beta_{2}=-\frac{1}{24}.\ee 
We note that in the expression above, there contains terms of the form $T J$ which follows directly from the Chern-Simons coupling.  The convention for the four dimensional Levi-Civita symbol is : $\epsilon^{txyw}=1.$

Recall that to linear order in $T$ and $J$ respectively the Chern-Simons term makes no contribution to the entanglement entropy.  Therefore the leading quadratic order correction from the extra terms in the entanglement entropy functional comes from the linear $J$ term in $\delta A$ coupled to the linear $T$ term in $h$. 

In general to compute second order change in the entropy one needs to evaluate the correction to the extremal surface.
Since the Chern Simons term vanishes in the AdS background and so does its corresponding entropy functional evaluated on the Rindler killing horizon, one can easily check that linear correction to the extremal surface $z_{1}$ does not enter in the leading contribution coming from the Chern Simons part of the entropy functional.  With the extremal surface remaining unchanged at (mixed) quadratic order, we can use (\ref{Waldentropy})  and (\ref{Waldnetropy1}) to compute the change in the entropy. The extremal surface sits on a constant time slice. The mixed terms would only be sensitivite to the $T_{ti}$ component, and so we will only keep those components. The change in the entropy coming from the Chern Simons part  is,
\be
\delta S_{EE, CS}=\delta S_{Wald, CS}
\ee
with the Killing vector,
\be
\zeta^{t}=\frac{2\pi}{R} (R^2-(t-t_{0})^2-(x-x_{0})^2-(y-y_{0})^2-(w-w_{0})^2)
\ee
and
\begin{align}
\begin{split}
\zeta^{z}=&-\frac{2\pi}{R}(t-t_{0}) z,\,\,\,\, \zeta^{x}=-\frac{2\pi}{R}(t-t_{0}) (x-x_{0}),\,\,\,\,\\ \zeta^{y}=&-\frac{2\pi}{R}(t-t_{0}) (y-y_{0}),\,\,\,\,\zeta^{w}=-\frac{2\pi}{R}(t-t_{0}) (w-w_{0}).
\end{split}
\end{align}
This gives,
\begin{align}
\begin{split} \label{a1}
\delta S_{EE, CS}=\frac{\lambda_{cs}}{\ell_{p}^3\,R}&\Big(24\,\pi \int_{\Sigma} d^3 x\, (R^2-r^2)^2\, \, x_{i}\,T_{t j}\, J_{k}\,\epsilon^{ijk}-8\,\pi \int_{\Sigma} d^3 x\, (R^2-r^2)^3\, \partial_{i}T_{t j}\, J_{k}\,\epsilon^{ijk}\\&+8\,\pi \int_{\Sigma} d^3 x\, (R^2-r^2)^2\, x_{i}\,x^{l}\partial_{l}T_{t j}\, J_{k}\,\epsilon^{ijk}\Big).
\end{split}
\end{align}
%where $\hat a, \hat b, \hat c$ are the three dimensional spatial indices. 
We have set for simplicity $x_{0}=y_{0}=w_{0}=0.$ Also we set the convention such that $\epsilon^{xyw}=1. $  After integrating by parts and noting that boundary terms evaluated at $R=r$ vanish due to the $R^2-r^2$ factors in the integrand, we have
\begin{align}
\begin{split}
\delta S_{EE, CS}=&\frac{8\,\pi\,\lambda_{cs}}{\ell_{p}^3\,R}\int_{\Sigma} d^3 x\,(R^2-r^2)(11r^2-7R^2)x_{i}T_{tj}J_{k}\epsilon^{ijk}\\&+\frac{8\,\pi\,\lambda_{cs}}{\ell_{p}^3\,R}\Big( \int_{\Sigma} d^3 x\, (R^2-r^2)^3\, T_{t j}\, \partial_{i}J_{k}\,\epsilon^{ijk}- \int_{\Sigma} d^3 x\, (R^2-r^2)^2\, x_{i}\,x^{l}T_{t j}\, \partial_{l}J_{k}\,\epsilon^{ijk}\Big).
\end{split}
\end{align}
This calculation can be checked against that obtained directly using the entropy functional derived in \cite{Gimseng}, whose explicit component form is detailed in the appendix. We find complete agreement.
Note that the  above expression contains the combination $x_i T_{tj} J_k \epsilon^{ijk}$ which appears like the coupling of angular momentum with the current. The integral is clearly vanishing for constant $T$ and $J$ since the integrand is odd. Therefore, we would keep up to linear order derivative terms in $T$ and $J$.

In addition to the CS term, the area functional descending from the Einstein-Maxwell terms would also contribute quadratic terms in $T$ and $J$, which would include these mixed $T.J$ terms. Terms that go like $T^2$ and $J^2$ have already been computed in \cite{Janet, Me}, which is shown to have coefficients with definite signs, and thus ensuring positivity of the relative entropy. In the following we will focus on the contribution from $T.J$ terms.

%To avoid clutter, we will assume that all the  $J_{i}$'s are spacetime independent functions, keeping explicit space-time dependence only in the stress tensor $T$.  
The contribution of $T.J$ coupled terms in the area functional could only arise from the $T.J$ terms appearing in the FG expansion of the metric $h$ in (\ref{pert1}), which is also the leading derivative correction. 
The correction to the area functional at this order is thus simply linear in $h$, 
\be
\delta S_{EE,Einstein} = \frac{2\pi \delta A }{\ell_{p}^3} =\frac{\pi R}{\ell_{p}^3} \int_{|x|\leq R} \frac{d^{3}x}{z_0^{4}} (h_i^i - h_{ij} \frac{x^ix^j}{R^2}) =\int_{\Sigma} \, d^{3}x\, \delta Q_{einstein}.
\ee
where $z_0 = \sqrt{R^2- x^2}$, which is the minimal entangling surface for a spherical region in  the pure AdS background.  

\begin{align}
\begin{split} 
\delta S_{EE,Einstein}=& \frac{2\pi R\, \lambda_{cs}\, }{\ell_{p}^3}\Big(\int_{\Sigma}\,d^{3}x\, (R^2-r^2)^2\,\partial_{i}T_{tj}J_{k}\epsilon^{ijk}+\int_{\Sigma} d^{3}x\, (R^2-r^2)^2\,\frac{x_{i}x^{l}}{R^2}\partial_{j}T_{tl} J_{k}\epsilon^{ijk}\Big)\\&
- \frac{3\pi R\, \lambda_{cs}\, }{\ell_{p}^3}\Big(\int_{\Sigma}\,d^{3}x\, (R^2-r^2)^2\,T_{tj}\partial_{i}J_{k}\epsilon^{ijk}+\int_{\Sigma} d^{3}x\, (R^2-r^2)^2\,\frac{x_{i}x^{l}}{R^2}T_{tl} \partial_{j}J_{k}\epsilon^{ijk}\Big).
\end{split}
\end{align}
After integrating by parts we get,
\begin{align}
\begin{split}
\delta S_{EE,Einstein}=&\frac{2\,\pi\,\lambda_{cs}}{\ell_{p}^3\,R}\int_{\Sigma}d^{3}x\, (R^2-r^2)(3R^2+r^2)x_{i}T_{tj}J_{k}\epsilon^{ijk}\\&- \frac{5\pi R\, \lambda_{cs}\, }{\ell_{p}^3}\Big(\int_{\Sigma}\,d^{3}x\, (R^2-r^2)^2\,T_{tj}\partial_{i}J_{k}\epsilon^{ijk}+\int_{\Sigma} d^{3}x\, (R^2-r^2)^2\,\frac{x_{i}x^{l}}{R^2}T_{tl} \partial_{j}J_{k}\epsilon^{ijk}\Big).
\end{split}
\end{align}

%-\frac{3}{2} \partial_{i}J_{j}T_{t\,k}\epsilon^{ijk} 
So, adding them,
\begin{align}
\begin{split} \label{mixed}
\delta_{\textrm{mixed}} S_{tot}=&\delta_{\textrm{mixed}} (S_{EE,Einstein}+ S_{EE,cs})\\&=\frac{10\,\pi\,\lambda_{cs}}{\ell_{p}^3\,R}\int_{\Sigma}d^{3}x\, (R^2-r^2)(9r^2-5R^2)x_{i}T_{tj}J_{k}\epsilon^{ijk}\\&+\frac{8\,\pi\,\lambda_{cs}}{\ell_{p}^3\,R}\Big( \int_{\Sigma} d^3 x\, (R^2-r^2)^2(\frac{3}{8}R^2-r^2)\, T_{t j}\, \partial_{i}J_{k}\,\epsilon^{ijk}- \int_{\Sigma} d^3 x\, (R^2-r^2)^2\, x_{i}\,x^{l}T_{t j}\, \partial_{l}J_{k}\,\epsilon^{ijk}\Big)\\&- \frac{5\pi R\, \lambda_{cs}\, }{\ell_{p}^3}\Big(\int_{\Sigma} d^{3}x\, (R^2-r^2)^2\,\frac{x_{i}x^{l}}{R^2}T_{tl} \partial_{j}J_{k}\epsilon^{ijk}\Big).
\end{split}
\end{align}
We would like to understand what these mixed contributions mean. 
Without loss of generality, we will only turn on $T_{ty}$ and $J_{w}$ component of the stress tensor and current.  Expressing the stress tensor in Fourier modes, we substitute
\be
T_{ty}=\alpha_{1}\,\cos(k_{1} x)+\beta_{1}\, \sin(k_{1} x)
\ee
and 
\be J_{w}=\alpha_{2}\,\cos(k_{2} x)+\beta_{2}\, \sin(k_{2} x). 
\ee
These choices are made to ensure that both $T$ and $J$ are conserved, and that $T$ is traceless. 

%In our convention ,
%\be
%Q_{einstein}=-\frac{1}{2\ell_{p}^3}\nabla^{a}\xi^{b}\epsilon_{ab}
%\ee
%and

Then we expand the result in small $k_{1}$ and $k_2$ and keep only up to terms linear in momenta.  
Spectacularly, at this order
\be \label{mixed2}
\delta_{\textrm{mixed}} S_{tot}=0,
\ee
even though the integrand is non-trivial, and begin to contribute in fact at order $k^3$. It is not clear to us whether this is an accident, or that such terms would be canceled out order by order in derivatives. If ever it does not cancel however, it is evident that they would violate unitarity, since it is also clear that they could not possibly combine with the $T^2$ and $J^2$ terms to form a perfect square. Recall from \cite{Janet,Me} that those terms to leading order in the derivative expansion do not depend on the CS coupling $\lambda_{cs}$. 

Note also that in this paper, we have considered a spherical entangling surface. It is known that the entanglement Hamiltonian of half-space  also admits a simple and local form. The spherical entangling surface is in fact related to the half-space entangling surface by a conformal transformation \cite{circle4}.  We note that the cancellation we observe here is by no means obvious at the level of the integrand, and the half-space calculation, being related to the current case by a conformal transformation, does not offer obvious new insights to the cancellation, although it will require extra complication and care to regulate the volume divergence of half-space that clutters computations here. 

%\be \label{mixed2}
%\delta_{\textrm{mixed}} S_{tot}=\frac{1024\,\pi^2 R^{8}\,\lambda_{cs}}{315\,\ell_{p}^3} \beta_{1}\,\alpha\,k_{1}.
%\ee
%{\bf For Janet: Following paragraph  should be modified}

%This should be compared with the quadratic contributions $T^2$ and $J^2$ terms.  The $T^2$ and $J^2$ contribution come only from the Einstein part which has been computed in \cite{Janet,Me}. 
%This is the leading contribution. From (\ref{mixed2}) and (\ref{mixed1}) it is evident that it is even if we add thes etwo terms there is no chance of forming a perfect square. Only (\ref{mixed1}) has a definite sign but not (\ref{mixed2}).
%[[For completeness we should include the $T^2$ and $J^2$ result.. just quote the previous papers in relative entropy paper of mine]]
%These do not involve the CS coupling $\lambda_{cs}$. As a result, for any small $k_1$, as soon as $\lambda_{cs} \sim (R k_1)^{-1}$, the $T.J$ term could potentially overwhelms the other quadratic contributions. This term however does not carry a definite sign, and that $\Delta \langle H \rangle$ carries only linear order contribution in $T$. Thus the positivity of the relative entropy between this excited state and the ground state 
%$ \Delta \langle H \rangle - \Delta S $ could turn negative at quadratic order in $T$ and $J$, violating unitarity. This suggests the following. 

One might wonder, whether it is possible that the entropy functional determined using the cone-method is subjected to ambiguity, and that extra terms could be added to cancel out these contributions should they contribute at higher order, and thus ensuring explicitly a positive definite relative entropy.

In the following, we will explore the possibility of extra terms in the entropy functional. 

 % It can be easily seen that there is no analogous contribution from the Einstein term.  So basically the first second order contribution to the total entanglement entropy is,
%\be
%\delta^{(2)} S_{EE}= \frac{24\,\pi\,\lambda_{cs}}{R} \int_{\Sigma} d^3 x (R^2-r^2)^2\, T_{t \hat a} (x-x_{0})_{\hat b}\, J_{\hat c}\,\epsilon^{\hat a \hat b \hat c}.
%\ee

\section{Modification of entropy functional ?} \label{sec4}

Now we turn our attention to the possibility of modifying the entropy functional such that these $T.J$ coupled terms can be canceled out explicitly. 

If any such term exist,  it is expected to satisfy the following criteria:
\begin{itemize}
\item It should not make any contribution on the killing horizon.
\item The leading contribution of which is a $T.J$ term. It should not alter the $T^2$ and $J^2$ terms. 
\item It is a covariant term constructed from the gauge field, the intrinsic and extrinsic curvatures on the entangling surface, and tangent and normal vectors thereof. 
\item It should carry at least 2-derivatives, as it was descended from the Chern-Simons action which carries four derivatives.
\item Since it is connected to the CS term, it carries an epsilon tensor, breaking time-reversal. 
\end{itemize}

We could come up with two such possibilities with exactly two derivatives:
\be
S_{extra,CS} = c_1 \int_{\Sigma} d^3x \epsilon^{ijk} A_i n^{\hat m\,\, l}\hat\nabla_j K^{\hat m}_{kl}  + c_2 \int_{\Sigma} A \wedge h^{\rho}_{\alpha}h^{\sigma}_{\beta}n^{\gamma}_{\tau}n^{\delta \kappa}R^{\tau}_{\kappa\rho\sigma}.
\ee
Here $\hat \nabla_i = P^{j}_i \nabla_j$, where  $P_{ij} =h_{ab}e^{a}_{i} e^{b}_{j}$ where, $e^{a}_{i}$ denotes the tangent vector.  $h_{ab}= g_{ab} - n^{\hat m}_a n^{\hat m}_b$ is the projector where  $n^{\hat m \in \{1,2\}}_i$ denotes the two normal vectors of the bulk entangling surface ($\Sigma$). 
 The first term vanishes identically because it is equivalent to
\be
\int_{\Sigma} d^3 x  \epsilon^{ijk} A_i K^{\hat m}_{jl}K^{\hat m\,\,l}_{k}.
\ee
The second term is actually related to 
\be
\int_{\Sigma} F\wedge\Gamma_N
\ee
as explained in the appendix.
This term, which is already present in the entropy functional following from the cone-method \cite{Gimseng, Dong, Dong1,Me1} in particular is controlled by anomaly. In the presence of the CS term in the bulk, the dual theory suffers from a mixed gauge-gravitational anomaly.  For example, under the  $U(1)$ gauge transformation $A\to A + d \Lambda$, we should recover 
% [[[[I am replacing $dU$ in Wall's paper by this term $\mathcal{F}$ defined in the appendix. Please correct it if they are not exactly equal]]]]
\be
\delta_\Lambda S_{EE} = 8\pi c_m \int_{\partial \mathcal{M}} \mathcal{F} \Lambda,
\ee  
where $\mathcal{F}$ is defined in the appendix (\ref{TP}), and $c_m$ is the anomaly coefficient related to the CS coupling by  \footnote{Invoking supersymmetry one can relate $\lambda_{cs}$ to $|\frac{a-c}{c}|$ as demonstrated in \cite{Cremonini, Aharony}, where $a$ and $c$ are Euler and Weyl anomaly coefficients in four dimensions.} 
%Also  this T.J correction is consistent with the local entropy increase theorem as demonstrated in 
%[[[Here what i am doing is to use eqn (\ref{Surface}), adjust it so that half of it is of the form $dA U - A dU$. The $A dU$ part controls the $U(1)$ transformation. Then compare coefficient ]]]
\be
c_m = \frac{\lambda_{cs}}{\ell_{p}^3}.
\ee
%\be
%c_m = \frac{2\pi}{8\pi \ell} (4 \lambda_{cs}/2).
%\ee

This anomaly coefficient is completely fixed by the theory, and so we are not entitled to add any extra term that changes this variation of the entanglement entropy. We have not been able to come up with any terms that are invariant both under diffeomorphism and $U(1)$ gauge transformation while satisfying all the requirements listed above. 

Let us also note that $A$ is proportional to $z^2 J$ and $\Gamma_{N} $ goes like $T$ to leading order in the FG expansion near the AdS boundary. Therefore, surface terms obtained from integrating by parts of the entropy functional (\ref{Surface}) when evaluated at the AdS boundary vanishes.  For that matter, surface terms carrying the $\epsilon$ tends to vanish at the boundary. 

To conclude, we believe that no additional terms, including surface terms, can be added to the entropy functional to remove the $T.J$ contribution explicitly.

\section{Time delay and causality constraint} \label{sec5}
%{\bf For Janet: Following paragraph should be modified}

The $T.J$ coupling was lying dangerously close to violating unitarity for any finite value of $\lambda_{cs}$. It prompted us to explore further consistency tests. There is a separate causality test that we would like to inspect in this section. It is well known that matter, photons or gravitons propagating through a gravitating background acquires a time-delay. Such a time-delay can be understood in terms of three point vertices, in which the propagating field is scattered off a graviton. In the presence of higher-derivative terms, the three point vertex could be corrected such that a time-delay arising purely from the Einstein action could be shifted into a time-advance, in which case causality from the point of view of asymptotic observers can be violated  \cite{Edelstein}.  
In the presence of the CS term, something rather interesting happens. The CS term provides a three point vertex $hhA$, in which a propagating graviton can be scattered off a photon. Therefore, it would be curious to see if a time-advance can be resulted from such couplings if we have a non-trivial charge background. 

To inspect such a possibility, we obtain the linearized equation of motion for the Einstein-Mixed Chern Simons theory around flat space in transverse traceless gauge in the presence of the background guage field. Below we quote the two results for linearized Riemann and Ricci tensor in a flat background with vanishing Christofel symbols,
\be
R^g{}_{bef}=\frac{1}{2}\Big(h^{g}{}_{f,be}-h_{bf}{}^{,g}_{,e}+h_{be}{}^{,g}_{,f}-h^{g}{}_{e,bf}\Big)
\ee
and
\be
R_{fb}=\frac{1}{2}\Big(-\Box h_{bf}\Big).
\ee
Choosing a transverse traceless gauge, the equations reduce to
\be
-\frac{1}{2}\Box h_{ab}=2\lambda_{cs}\Big(\epsilon^{cdef}{}_{(a} h^{g}{}_{f,b)e}\nabla_{g}F_{cd}-\epsilon^{cdef}{}_{(a}h_{b)f}{}^{,g}_{,e}\nabla_{g}F_{cd}- \epsilon^{cdef}{}_{(a}F_{cd}\nabla_{e}\Box h_{fb)}\Big).
\ee
We will work in a flat background, in which we choose coordinates such that the background metric is
\be \label{metric2}
ds^2=-2 du dv+\delta _{ij}dx_\perp^{i} dx_\perp^{j}.
\ee
where $i,j$ are the spatial indices. We will also take $h_{ab}=\epsilon_{ab} f(u,v,x_\perp) $ where $\epsilon_{ab}$ is the polarization tensor, and $x_\perp$ denotes the $d-2$ directions $x_\perp^i$. 
Now, consider a charged background in which
\be
F_{ui} = \sigma \frac{\theta(u)}{|r_\perp|} \hat{r}_{\perp\,\, i},
\ee
where $\hat{r}_{\perp}^i = x^i_\perp/|r_{\perp}| $ and $\sigma$ is some constant.
This background can be understood to be a solution of Maxwell equations in flat space with a simple charge density $\mathcal{J}$, 
\be
\partial^\alpha F_{\alpha u} = \partial_i F_{i u} = \mathcal{J}_u,\qquad \mathcal{J}_u = \sigma \frac{(d-4)\theta(u)}{|r_\perp|^2},
\ee
while all other components of $\mathcal{J}$ vanishes. This current density is conserved, satisfying
\be
-\frac{\partial_v \mathcal{J}_u }{2}+ \partial_i \mathcal{J}_i = 0.
\ee
Note that,
\be
\partial_u F_{uj} = \sigma\frac{\delta(u)}{|r_\perp|}\hat{r}_{\perp\,\, j}, \qquad \partial_i F_{uj} = \sigma \frac{\theta(u)}{|r_\perp|^2}  (\delta_{ij} - 2\hat{r}_{\perp\,\,i}\hat{r}_{\perp\,\, j}).
\ee

We note that when $|r_\perp|$ is large, i.e. large impact parameter, $\nabla_{u}F_{ud}$ is dominant. 
We would also take the large momentum limit, in which $\partial_{u,v} h \gg \partial_{i}h$. 
%Using all these approximation we get the following linearized equation for graviton, 
%\be
%\Big(-\frac{1}{2}\delta^{f}_{(a} +2\lambda_{cs}\,\epsilon^{cdef}{}_{(a}F_{cd}\nabla_{e}\Big)\Box h_{fb)}=2\lambda_{cs}(\epsilon^{cdef}{}_{(a} h^{g}{}_{f,b)e}\nabla_{g}F_{cd}-\epsilon^{cdef}{}_{(a}h_{b)f}{}^{,g}_{,e}\nabla_{g}F_{cd}\Big)
%\ee
%Now we make  further the following assumptions,
%$h_{ui}=h_{vi}=0$ and $\nabla_{v}F_{ui}$ is  dominant over $F_{\mu\nu}.$  

Now we replace
\be
h_{ab} = \epsilon_{ab} f(u,v,r_\perp),
\ee
where $\epsilon_{ab}$ denotes the polarization tensor.
Contract the equation of motion by $\epsilon^{ab}$, making the above approximations we get, 
\be 
\epsilon^{ab}\epsilon_{ab} \partial_{u}\partial_{v} f=\lambda_{cs}\epsilon^{udvf}{}_{a}\epsilon^{u}_{f}\epsilon^{av}\nabla_{u}F_{ud}\partial_{v}^2 f. 
\ee
Note that  $g^{uv}=-1$.
%The right hand side is non vanishing only when $f$ belongs to the three dimensional spatial indices.  We denote it by $l.$ So finally we get,
%\be
%-\frac{1}{2}\Box h_{ij}=-2\lambda_{cs}\epsilon^{ukvl}{}_{(i} h_{j) l,v}{}^{v}\nabla_{v}F_{uk}
%\ee
This takes precisely the same form as the computation in which a graviton is propagating in a shock wave background \cite{Edelstein}, leading to a time delay given by
\be\label{timedelay}
\Delta v = \frac{ - \lambda_{cs}  \epsilon^{uivfj}\epsilon^u_f\epsilon^{jv} \hat{r}_{\perp\,\, i} }{\epsilon_{ab}\epsilon^{ab}}  \frac{\sigma}{|r_\perp|}.
\ee
This would suggest that $\lambda_{cs}$ should remain perturbatively small to preserve asymptotic causality, for the same reason put forward in \cite{Edelstein}.

%Now we expand the box using the metric mentioned  (\ref{metric2}) and we will  drop the spatial dependence from the graviton fluctuation. But first we write the eom in the following manner, 
%\be
%\delta_{il} \Box h_{jl}-4 \lambda_{cs} \epsilon^{ukvl}{}_{(i} h_{j) l,v}{}^{v}\nabla_{v}F_{uk}=0.
%\ee
%Then further expanding it we get,
%\be
%\delta_{il}g^{uv}\partial_{u}\partial_{v}h_{jl}+\Big(\delta_{il}-4\lambda_{cs}\epsilon^{ukvl}{}_{(i}\nabla_{v}F_{uk}\Big) g^{vv} \partial_{v}^2h_{j) l}=0.
%\ee

\section{Discussion} 
%{\bf For Janet: Following lines should be modified}

In this note we have looked into the relative entropy of a holographic theory in the presence of Chern Simons terms. Perturbing around the pure AdS background by turning on the metric and gauge fluctuations sourced by a small stress tensor and current density in the boundary theory, we find that the entanglement entropy acquires a $T.J$ correction that could potentially violate the positivity of the relative entropy for any finite value of the CS coupling. We note that such a contribution is only present at exactly $d=5$. For higher dimensional CS theories, the leading mixed contributions would appear at least quadratic order in $T$, and thus manifestly subleading to $T^2$ contributions, leading to no constraints.   While it was not possible to cancel these mixed terms by adding extra terms to the entropy functional without contradicting the correct transformation of the entropy controlled by anomaly,
miraculously however, at least to linear derivative order, these terms cancel out. 

 We complement our analysis by inspecting time delay of gravitons propagating in a charged background. Analogously to other higher derivative theories, there is the possibility of  violating asymptotic causality at any finite values of the CS coupling.  
 This $T.J$ correction is also consistent with the positivity of the entropy current which also establishes a local entropy increase theorem for this Chern-Simons theory as demonstrated in \cite{Chapman} upto certain order in perturbation theory. It will be an interesting future direction to get the perturbative nature of $\lambda_{cs}$ from the second law and establish GSL for this theory following \cite{Me3, Wall1}. In higher dimensions, the relative entropy would not give any useful constraints to the couplings of Chern-Simons terms, for reasons mentioned above. It will be interesting to try to constrain those couplings using other methods, for example  using the black hole second law.

%[[[Please add anything you find appropriate, such as any comments you might have about GSL etc etc]]]

% \be
%G^{\mu\nu}=2\lambda_{cs} \epsilon^{\rho\sigma\tau \lambda (\mu}\nabla_{\alpha}(F_{\rho\sigma} R_{\tau \lambda}{}^{\nu)\alpha})
%\ee
\section*{Acknowledgements} 
AB, LC and LYH would like to acknowledge support by the Thousand Young Talents Program, and Fudan University. LC is also partially supported by  China postdoctoral Science Foundation with Grant No.2016M591593.

%A acknowledge B and C for.... 
\appendix
\section{Wald charge and its variation $\delta Q$}

We explain the computation of the Wald charge here following \cite{Thesis} \footnote{We have also corrected several typos found in \cite{Thesis}.}.

Starting from the action (\ref{CSL}),  we shall first perform variation of the charged Lagrangian evaluating the variations of gravitational and gauge fields, which takes the form as
\bea
\D (L_\textrm{maxwell}+L_\textrm{CS})=d\Theta+(E^g)^{ab}\D g_{ab}+(E^A)^{a}\D A_{a}
\eea
where \footnote{We will reinstate the overall factor of $\frac{1}{2\,\ell_{p}^3}$ at the end.},
\bea
L_\textrm{maxwell}+L_\textrm{CS}=\Big(-\frac{1}{2}F\wedge*F+\lambda_{cs} \tr(A\WE R\WE R)\Big)
\eea
is the charged Lagrangian. The equation of motion form  for the metric and the $U(1)$ gauge field are
\bea
&&(E^g)^{ab}=-\frac{\lambda_{cs}}{4}\ep^{gacde}\de_f(F_{cg}R_{de}^{~~fb})-\frac{\lambda_{cs}}{4}\ep^{gbcde}\de_f(F_{cg}R_{de}^{~~fa})+\frac{g_{ab}}{8}F^2-\frac{1}{2}F_{af}F^b_{~f};\nonumber\\
&&(E^A)^{a}=\de_bF^{ba}+\lambda_{cs} \frac{\ep^{abcde}}{4}R_{bcfg}R_{de}^{~~fg},
\eea
the boundary term is
\bea
\Theta=-\D A\WE*F+\lambda_{cs} \left(-\ep^{v abde}A_b\de_f\D g_{ac}R_{de}^{~~fc}+\frac{\ep^{abcde}}{2}F_{ac}R_{de}^{~~v f}\D g_{bf}\right){\bep_v}.
\eea

So the Noether current is given by
\bea
J=\Theta-i_{\xi}L
\eea
with
\bea
i_{\xi}(L_\textrm{maxwell}+L_\textrm{CS})&=&-\frac{1}{2}\left(i_{\xi}F\WE*F+F\WE i_{\xi}*F\right)+\lambda_{cs}\,\tr\left(i_{\xi}A\WE R\WE R-A\WE i_{\xi}R\WE R-A\WE R\WE i_{\xi}R\right)\nonumber\\
&=&\left(-\frac{1}{4}F^2\xi^v+\frac{\lambda_{cs}}{4}\ep^{v bcde}\xi^{a}A_a R_{bcfg}R_{de}^{~~fg}-\lambda_{cs}\,\ep^{v acde}A_a\xi^bR_{bcfg}R_{dc}^{~~fg}\right){\bep_v}.
\eea

From the Noether current, a tedious but straightforward calculation deduces  the conserved charge
\bea 
J&=&dQ+2\lambda_{cs}\,(E^g)^{ab}\xi_b+(E^A)^{a}\xi^bA_b\nonumber\\
&=&\de_g\left[-\lambda_{cs}\Big(\ep^{a gbde}A_bR_{de}^{~~fv}\de_f\xi_v-\frac{\ep^{a fbde}}{2}F_{fb}R_{de}^{~~gv}\xi_v+\frac{\ep^{gbcde}}{2}F_{bc}R_{de}^{~~a f}\xi_f-\frac{\ep^{fbcde}}{2}F_{fc}R_{de}^{~~a g}\xi_b\Big)-F^{a g}\xi^bA_b\right]\nonumber\\
&&~~~~~~~~~~~~~~~~~~~~~~~~+2\lambda_{cs}\,(E^g)^{ab}\xi_b+(E^A)^{a}\xi^bA_b,
\eea
so the conserved charge $Q$ is
\bea
Q=\frac{1}{2}Q^{ag}\bep_{ag}
\eea
with  
\bea   \label{cs1}
Q^{ag}_{CS}=\ep^{a gbde}A_bR_{de}^{~~fv}\de_f\xi_v-\frac{\ep^{a fbde}}{2}F_{fb}R_{de}^{~~gv}\xi_v+\frac{\ep^{gbcde}}{2}F_{bc}R_{de}^{~~a f}\xi_f-\frac{\ep^{fbcde}}{2}F_{fc}R_{de}^{~~a g}\xi_b\nonumber\\
\eea
 and
 \bea
 Q_{maxwell}^{ag}= F^{a g}\xi^bA_b.
 \eea
After adding the Einstein piece we get the full expression for the charge in Einstein-Maxwell-Chern Simons theory.
\be \label{totcharge}
Q_{tot}^{ag}= \frac{1}{2\ell_{p}^3}\Big(Q_{EH}^{ag}+Q_{maxwell}^{ag}+\lambda_{cs} Q_{CS}^{ag}\Big).
\ee
So,
\be
Q_{tot}= \frac{1}{2}Q^{ag}\epsilon_{ag}.
\ee

\section{Entropy functional}

The total entropy functional is given by 
%[[[[[please check this relative $2\pi$???]]]]]
\be\label{gmsg}
S_{EE}=  \frac{2\pi }{\ell_{p}^3}\int_{\Sigma} d^{3} x\Big( \sqrt{h} +8 \lambda_{cs} \, F \wedge \Gamma_{N}\Big),
\ee
where $\Gamma_{N}$ is the twist potential and $\Sigma$ denotes the entangling surface.  Now after doing an integration by part we can write this,
% [[same issue about this relative factor of $2\pi$.]]]
\be \label{Surface}
S_{EE}= \frac{2\pi }{\ell_{p}^3}\int_{\Sigma} d^{3} x\Big[ \sqrt{h} +8 \lambda_{cs} \,(d ( A \wedge \Gamma_{N})-A\wedge d\Gamma_{N})\Big],
\ee
where, $F=dA.$ Now we can use the following identity known as the Voss-Ricci relation, which is related to the derivative of the twist potential of the extrinsic curvature of the surface and  Riemann tensor \cite{Iqbal}. 
\be \label{VR}
\mathcal{F}_{ab}{}^{cd}=\Big(\mathcal{K}^{c}{}_{ae}\mathcal{K}^{d}{}_{bf}-\mathcal{K}^{c}{}_{be}\mathcal{K}^{d}{}_{af}\Big)g^{ef}+h^{e}_{a}h^{f}_{b}n^{c}_{h}n^{d\, g}R^{h}_{}{g e f},
\ee
where, $\mathcal{F}_{ab}{}^{cd}=\mathcal{F}_{ij}^{\hat m \hat n}e^{i}_{a}e^{j}_{b}n^{\hat m\,c}n^{\hat n \,d}.$ The field strength for the twist poetntial is defined below,
\be \label{TP}
\mathcal{F}_{ij}^{\hat m \hat n}\equiv\partial_{i}\Gamma^{\hat m \hat n}_{Nj}-\partial_{j}\Gamma^{\hat n \hat m }_{N i}.
\ee
Also the covariant extrinsic curvature is defined by, $\mathcal{K}^{c}_{ab}\equiv\mathcal{K}^{\hat m}_{ij}n^{\hat m\,c}e^{i}_{a}e^{j}_{b}$ where $\mathcal{K}^{\hat m}_{ij}$ is the extrinsic curvature defined on the codimension-2 extremal surface.  We define the induce metric on the surface, $h_{ij}=e^{a}_{i}e^{b}_{j}g_{ab}$ where $g_{ab}$ is the bulk metric and  $n_{ab}=n^{\hat m}_{a}n^{\hat n}_{b}\delta_{\hat m \hat n }.$ Now we know that $d\Gamma_{N}=\mathcal{F}.$ Then $A\wedge d \Gamma_{N}=\frac{1}{2}A_{i}\mathcal{F}_{jk}\epsilon^{ijk}.$ Further we can use $\mathcal{F}_{ij}=\frac{1}{2}\epsilon^{\hat n\hat m}e^{a}_{i}e^{b}_{j}\mathcal{F}_{ab}{}^{ \hat n \hat m}=\frac{1}{2}\hat \epsilon^{cd} e^{a}_{i}e^{b}_{j}\mathcal{F}_{abcd}$ where $\hat \epsilon^{cd}= \epsilon^{\hat n \hat m}n^{\hat n\,c}n^{\hat m\,d}$ is the bi-normal defined on the entangling surface. Then if one use the relation defined in (\ref{VR}), one can easily check that the entropy functional mentioned in (\ref{gmsg}) is equivalent to the entropy functional obtained by using the formula in \cite{Dong} upto a surface term as mentioned in (\ref{Surface}).\\
%[[[Maybe too sketchy above... can you add a bit more details??
Focussing on the Chern-Simons part of the action (upto an overall factor) , we express the entropy functional in explicit component forms to facilitate computations. 

\be
S_{EE, CS}=\int_{\Sigma} d^{3} x \epsilon^{ijk} F_{ij} \Gamma_{N\, k}.
\ee
 Now we have to write the $\Gamma_{N}$ which the connection for the normal bundle  in term of the metric to make further computation \cite{Me1},
\be
\Gamma_{ N\, k} =\epsilon^{\hat n \hat m}\delta_{\hat r\hat n}\Gamma^{\hat r}_{k \hat m}=\epsilon^{\hat n\hat m}\delta_{\hat r \hat n}(\partial_{k }n^{\hat m\,a}+\tilde \Gamma^{a}_{bc}e^{b}_{k}n^{\hat m\,c})n^{\hat r }_{a},
\ee
$\tilde \Gamma^{a}_{bc}$ is the bulk connection. $\hat m, \hat n, \hat r$ are indices denoting the transverse direction to the entangling surface which is in our case $z$ and $t$. Also,
\be
n_{\hat r}.n_{\hat m}=\pm \delta_{\hat r \hat m}.
\ee
$\epsilon^{\hat m \hat n }$ is the two dimensional Levi-Civita tensor in the transverse space, $n^{a}$ is the normal defined for the surface and $e^{b}_{k}$ is the tangent vector.

\end{document}